\documentclass[twocolumn,aps,prl,showpacs,byrevtex]{revtex4-1} %superscriptaddress,
\usepackage{epsf,graphicx,amssymb}
\usepackage[usenames]{color}
\usepackage{natbib}
\usepackage{float,ulem}
\usepackage{upgreek}
\usepackage{siunitx}
\usepackage{gensymb}
\usepackage{amsmath,amssymb}
\usepackage{color}
\begin{document}

\title{Statistical Equilibrium of Large Scales in three-dimensional Hydrodynamic Turbulence}
%Experimental evidence of statistical equilibrium of large scales\\ in three-dimensional hydrodynamic turbulence
%Statistical equilibrium of large scales (experimentally achieved) in three-dimensional hydrodynamic turbulence

\author{Jean-Baptiste Gorce}
\affiliation{Universit\'e Paris Cit\'e, CNRS, MSC Laboratory, UMR 7057, F-75 013 Paris, France}
%\affiliation{Universit\'e de Paris, MSC Laboratory, UMR 7057 CNRS, F-75 013 Paris, France}
\author{Eric Falcon}
\email[E-mail: ]{eric.falcon@u-paris.fr}
%\href{https://orcid.org/0000-0003-4221-7622}{\includegraphics[width=1]{ORCIDiD_iconvector.svg}
%\orcid{0000-0001-9640-9895}
\affiliation{Universit\'e Paris Cit\'e, CNRS, MSC Laboratory, UMR 7057, F-75 013 Paris, France}

%\date{\today}
\begin{abstract}
We investigate experimentally three-dimensional (3D) hydrodynamic turbulence at scales larger than the forcing scale. We manage to perform a scale separation between the forcing scale and the container size by injecting energy into the fluid using centimetric magnetic particles. We measure the statistics of the fluid velocity field at scales larger than the forcing scale (energy spectra, velocity distributions, and energy flux spectrum). In particular, we show that the large-scale dynamics are in statistical equilibrium and can be described with an effective temperature, although not isolated from the turbulent Kolmogorov cascade. In the large-scale domain, the energy flux is zero on average but exhibits intense temporal fluctuations. Our work paves the way to use equilibrium statistical mechanics to describe the large-scale properties of 3D turbulent flows.
\end{abstract}

\maketitle

\paragraph*{Introduction.\textemdash}Three-dimensional (3D) hydrodynamics turbulence has been extensively studied to characterize the energy transfers in the inertial range, the interval between the energy injection scale and the small (dissipative) scale~\cite{Richardson,Kolmogorov41,K41b,Frisch,Pope}. While they control many properties of 3D turbulent flows, e.g., mixing in industrial flows, or transport of tracers in geophysical and astrophysical turbulent flows~\cite{Moffatt}, the large-scale properties of turbulence, the scales larger than the forcing scale, have been less investigated. Indeed, in most experiments and direct numerical simulations (DNS), 3D turbulent flows are forced at a scale comparable to the container size to study the turbulent energy cascade within the inertial range. However, it has been conjectured that the large-scale modes of turbulent flows possess the same energy and are in a statistical stationary equilibrium regime~\cite{Burgers29,Frisch,Hopf,Lee,Kraichnan}. This equipartition regime, also called thermal equilibrium, would occur if no mean energy flux is transferred from the forcing scale to the large scales. Such a statistical equilibrium is difficult to observe in most experimental systems and numerical simulations because there is no scale separation between the forcing scale and the container size. Numerical simulations have recently confirmed the statistical equilibrium in 3D forced turbulent flows for the spectrally truncated Navier-Stokes equations~\cite{AlexakisJFM2020} and the truncated Euler equation~\cite{Cichowlas2005,Dallas2015,AlexakisJFM2019,Verma2020}, but experimental evidence of this regime remains elusive.

Here, we generate 3D hydrodynamic turbulence using centimetric magnetic particles immersed in a large fluid reservoir. This method provides a wide interval between the energy injection scale and the container size. We observe a statistical equilibrium regime in this large-scale interval while a turbulent cascade develops in the inertial range. We also show that the effective temperature of the statistical equilibrium regime is related to the injection of energy. Note that large-scale structures in decaying 3D turbulence have been investigated~\cite{Batchelor56,Saffman67,Davidson,Lesieur,DavidsonJFM2012,YoshimatsuPRF2019}, but are different from the stationary (forcing) case~\cite{AlexakisJFM2019}. Other turbulent systems also exhibit large-scale statistical equilibrium, e.g., in wave turbulence with no inverse cascade, such as capillary waves~\cite{Balkovsky1995,Michel2017,Abdurakhimov2015}, bending waves in mechanical plates~\cite{MiquelPRE2021}, or optical waves~\cite{Josserand05}. Conversely, the presence of an inverse cascade implies that two-dimensional hydrodynamic turbulence~\cite{Kraichnan67}, gravity wave turbulence~\cite{FalconPRL2020}, or acoustic wave turbulence in superfluid~\cite{Ganshin2008} do not exhibit a statistical equilibrium regime.

\paragraph*{Theoretical backgrounds.\textemdash} In the case of nonhelical, incompressible, inviscid, and force-free turbulent flows, Kraichnan~\cite{Kraichnan} derived the statistical equilibrium energy spectrum $E_T(k)$ for low wavenumbers $k$
\begin{equation}
E_{T}(k)=\frac{4\pi \alpha k^2}{\alpha^2-\beta^2k^2}.
\label{kraichnan1}
\end{equation}
$\alpha$, $\beta$ are determined by the total energy and the helicity of the system. This result referred to as {\it absolute equilibrium}, is related to classical equilibrium statistical mechanics and is equivalent to the equipartition of the total kinetic energy among the large-scale Fourier modes~\cite{Rose1978,Lesieur,Hopf,Lee}. One can also say that the large-scale equipartition implies that the spectral energy density per unit mass $E_T(k)\mathrm d k$ % within a sphere of radius $k$ and thickness $\mathrm d k$ 
is equal to the number of modes times the energy per mode per mass, i.e.,
\begin{equation}
E_{T}(k)\mathrm d k=\frac{4\pi k_{\rm B}{T}}{\rho}k^2 \mathrm d k
\label{kraichnan}
\end{equation}
with $k_{\rm B}$ the Boltzmann constant, $\rho$ the fluid density, and $T$ a temperature in a classical thermodynamic equilibrium sense. Therefore, one obtains the expression which is equivalent to Eq.~\eqref{kraichnan1} when $\beta k \ll \alpha$. Note that deviations from Eq.~\eqref{kraichnan1} are expected for a broadband spectral forcing instead of a narrow one~\cite{AlexakisJFM2019} or in the case of anisotropic turbulence~\cite{Thalabard2015}. 

By assuming high-Reynolds number isotropic turbulence, the energy spectrum in the inertial range (i.e., for high $k$) is given by the Kolmogorov spectrum~\cite{Kolmogorov41} 
\begin{equation}
E_K(k)=C_K\epsilon^{2/3}k^{–5/3}
\label{kolmogorov}
\end{equation}
with $C_K\simeq 1.6$ the experimentally measured Kolmogorov constant \cite{SaddoughiJFM94}, and $\epsilon$ the rate of energy dissipation per unit mass. In the stationary regime, $\epsilon$ is constant and equal to the energy flux transferred from the forcing scale to the dissipation scale. Therefore, $\epsilon$ is also equal to the energy injection rate in the stationary regime.

\paragraph*{Experimental setup.\textemdash}We inject energy homogeneously into the fluid using small magnetic particles. The nonlinear transfer of energy and turbulent cascade towards small scales (inertial range) have been characterized using this method~\cite{CazaubielPRF2021}. To measure the large-scale properties of turbulence, we scaled up this experimental system. A plexiglass square container of length $L=32$\,cm and height $h=22$\,cm is filled with water (22.5\,L) and sealed by a transparent lid. This fluid container sits between a pair of Helmholtz coils (0.49\,m inner diameter and 1\,m outer diameter). The pair of coils is powered by a sinusoidal current (Itech IT7815 AC 15 kW power supply) and generates a vertical oscillating magnetic field $B(t)$ with an amplitude $B\in[0, 360]$\,G and a frequency $F\in[0, 25]$\,Hz. This AC magnetic field is homogeneous within all the volume of the fluid container (5\% accuracy) and transfers kinetic energy to $N$ neodymium magnets encapsulated in plexiglass shells (1 cm), which are immersed in the fluid ($N \in[50, 450]$). The volume fraction  of the magnetic particles is smaller than 1.5\%. The kinetic energy of the magnetic particles is then transferred to the surrounding fluid randomly in both space and time (see~\cite{FalconEPL2013,FalconPRF2017,CazaubielPRF2021} and movies in the Supplemental Material~\cite{SuppMat,Note}).  The forcing scale is estimated to be 5\,cm. It corresponds to the integral scale $L_i$ defined as the abscissa of the maximum of the energy spectrum (see below). Note that $L_i$ cannot be accurately computed from the autocorrelation function of the velocity field since the container size $L$ is not eight times larger than the integral scale~\cite{Pope,CazaubielPRF2021,ONeill2004}. The fluid velocity is measured locally by nonintrusive Laser Doppler Velocimetry (LDV Dantec Flow Explorer 1D) with a sampling frequency of 250\,Hz. We perform Particle Image Velocimetry (PIV)~\cite{Thielicke2014} to measure the fluid velocity field in a horizontal $xy$ plane ($32\times 32$\,cm$^2$). The fluid is seeded with Polyamide fluid tracers (\SI{50}{\upmu m}) illuminated by a horizontal laser sheet and a high-speed camera (Phantom V1840, 2048\! $\times$\! 1952 pixels$^2$ at 200 fps), located on the top of the fluid container, records time series of images. The mean fluid velocity is smaller than the standard deviation of the velocity fluctuations $\sigma_u$ ($<10$\%), such that one can assume that there is no mean flow. The isotropy of the velocity field is also checked for different values of $N$~\cite{Note2}. Typical values of the turbulent flow are the following: the dissipation rate $\epsilon$ is $3 \times 10^{-4}$\,m$^2$/s$^3$, the Reynolds number at the integral scale $L_i\simeq 5$\,cm is 650, and the Reynolds number at the Taylor scale $L_{\lambda}\simeq 7.6$\,mm is ${\rm Re}_{\lambda}=100$. We have ${\rm Re}_{\lambda}\in[56,100]$ when changing the experimental parameters ($F$, $N$, or $B$).

\begin{figure}[!t]
\begin{center}
\includegraphics[width=8.6cm]{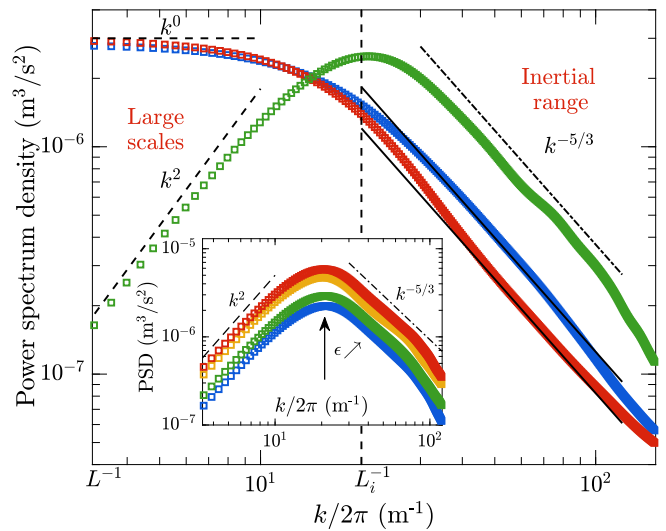} %fig2New.eps
\caption{3D power spectrum density $E(k)$ (green) derived from the 1D spectrum of the longitudinal velocity $E_{uu}(k_x)$ (red), and transverse velocity $E_{vv}(k_x)$ (blue), (Eq.~\eqref{3Dspectrum}). Dashed line: $k^2$ power law illustrating the large-scale statistical equilibrium regime. Dot-dashed line: $k^{-5/3}$ power law illustrating the inertial range of the turbulent cascade. The vertical dashed line corresponds to the inverse of the integral scale $k_i/2\pi=1/L_i$ and separates the large-scale domain ($k<k_i$) from the inertial range ($k>k_i$). The PIV measurements are performed at $F=20$\,Hz, $B=290$\,G and $N=55$. Inset: Power spectrum densities at different $\epsilon\in[1.1,3.2]\times10^{-4}$\,m$^2$/s$^{3}$.} 
\label{fig01}
\end{center}
\end{figure}

\paragraph*{Spatial power spectrum.\textemdash}The longitudinal and transverse horizontal fluid velocities are defined as $u(x,t)$ and $v(x,t)$. We first measure the longitudinal $E_{uu}(k_x)$ and transverse $E_{vv}(k_x)$ spectra  (Fig.~\ref{fig01}). The inertial range is consistent with the Kolmogorov prediction over a decade. The power spectra are proportional to $k_x^{-5/3}$ in the inertial range and the ratio between the unidimensional (1D) power spectra is equal to $E_{vv}(k_x)/E_{uu}(k_x)=4/3$~\cite{Pope} (black lines in Fig.~\ref{fig01}). In the case of isotropic turbulence, the 3D power spectrum $E(k)$ is derived from the longitudinal and transverse spectra~\cite{Pope,Davidson}
\begin{equation}
E(k)=-k\frac{\mathrm d}{\mathrm d k}\left[\frac{1}{2}E_{uu}(k_x)+E_{vv}(k_x)\right]\, {\rm .}
\label{3Dspectrum}
\end{equation}
The energy spectrum $E(k)$ is shown in Fig.~\ref{fig01}. A $k^2$ power law is observed in the energy spectra at lower wavenumbers, illustrating the statistical equilibrium regime, while a
$k^{-5/3}$ power law is observed at higher wavenumbers, indicating a direct energy cascade in the inertial range. In between these regimes, the wavenumbers close to the value $k_i=2\pi/L_i$ suggest that the statistical equilibrium state and the out-of-equilibrium one interact with each other (see below).

\begin{figure}[!t]
\begin{center}
\includegraphics[width=8.6cm]{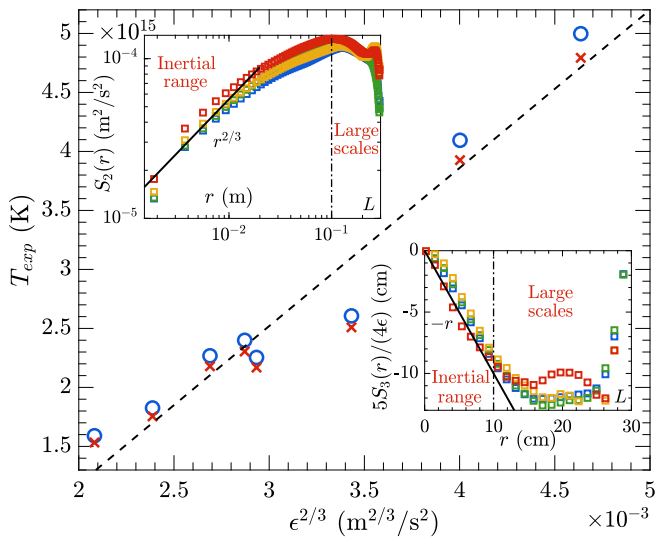}
\caption{Effective temperature $T_{exp}$ of the statistical equilibrium regime of the large scales for different energy injection rates $\epsilon$. The energy injection rate $\epsilon$ is measured using $\epsilon=2\nu \int_{2\pi/L}^{\infty} k^2 E(k)dk$. $T_{exp}$ is measured from the 3D spectra shown in Fig. \ref{fig01} using Eqs. \eqref{kraichnan} ($\circ$) and \eqref{Texp} ($\times$). The solid dashed line corresponds to Eq.~\eqref{balance} using $k_i/2\pi=13.3$ m\textsuperscript{-1}. Insets: Structure functions $S_2(r)$ (top) and $S_3(r)/(4\epsilon/5)$ (bottom) for different $\epsilon$ (same colors as in the inset of Fig.~\ref{fig01}). Solid lines correspond to $r^{2/3}$ and $-r$, respectively.}
\label{fig02}
\end{center}
\end{figure}

\paragraph*{Effective temperature.\textemdash} Both $k^{2}$ and $k^{-5/3}$ power laws are consistently observed in the energy spectra when increasing the energy injection rate $\epsilon$ (inset of Fig.~\ref{fig01}). For each 3D spectrum $E(k$), we compute the effective temperature by integrating both members of Eq.~\eqref{kraichnan} in the large-scale interval $k\in [k_L=2\pi/L$, $k_i=2\pi/L_i]$
\begin{equation}
T_{exp}=\frac{3\rho}{32\pi^4k_{\rm B}}\frac{L^3L_i^3}{L^3-L_i^3}\int_{k_L}^{k_i}E(k)\mathrm d k\, {\rm .}
\label{Texp}
\end{equation}
$T_{exp}$ is shown in Fig.~\ref{fig02} as a function of the energy injection rate $\epsilon$ ($\times$). One can also estimate the temperature by fitting directly the experimental 3D spectra with Eq.~\eqref{kraichnan}, which leads to similar values of the temperature ($\circ$ in Fig.~\ref{fig02}). $T_{exp}$ is found to be 13 orders of magnitude higher than the room temperature and proportional to $\epsilon^{2/3}$ (Fig.~\ref{fig02}). This result is explained by equating Eqs.~\eqref{kraichnan} and \eqref{kolmogorov}, which gives the relationship 
\begin{equation}
T=\frac{\rho C_K}{4\pi k_{\rm B}}k_i^{-11/3}\epsilon^{2/3}.
\label{balance}
\end{equation}

\paragraph*{Structure functions.\textemdash}The velocity increments at a distance $r$, $\mathcal{S}_i(r)=\langle[v(x+r)-v(x)]^i\rangle$ are now computed from the PIV measurements. The insets of Fig.~\ref{fig02} show that $\mathcal{S}_2(r)\sim (\epsilon r)^{2/3}$ and $\mathcal{S}_3(r)=-4\epsilon r/5$ in the inertial range, as predicted theoretically~\cite{Kolmogorov41,K41b,Pope}. For large scales (i.e., $r>0.1$\,m), $\mathcal{S}_2(r)$ and $\mathcal{S}_3(r)$ are found to be roughly independent of $r$, except when $r\simeq L$. We also found that $\mathcal{S}_2(r)\simeq 2 \sigma_u^2$, suggesting that the velocities are uncorrelated at long distances, as expected~\cite{Davidson}. 

\begin{figure}[!t]
\begin{center}
\includegraphics[width=8.6cm]{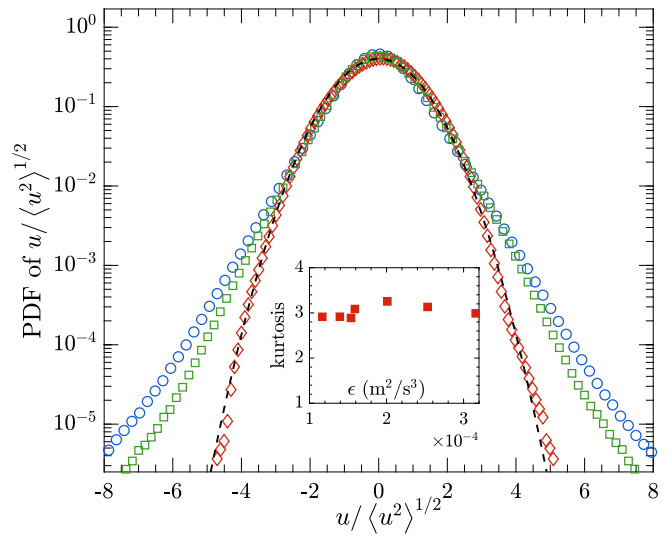}
\caption{Probability density functions (PDF) of the normalized fluid velocity fluctuations $u/\sqrt{\langle u^2\rangle}$ of ($\Diamond$) large-scale modes, ($\square$) all modes, and ($\circ$) small-scale modes for $\sigma_u=0.6$ cm/s. The cutoff value of the filter is equal to $k/2\pi=9.4$ (m\textsuperscript{-1}). The black dashed line represents a Gaussian distribution. Inset: Kurtosis ($K=\langle u^4\rangle/\langle u^2\rangle^2$) of the large-scale modes as a function of $\epsilon$ ($\sigma_u\in[0.6,1.3]$ cm/s).} 
\label{fig03}
\end{center}
\end{figure}

\paragraph*{Velocity probability distribution.\textemdash}
The probability distribution functions (PDF) of the velocity field (Fig.~\ref{fig03}) are found to be strongly non-Gaussian probably because the PDFs of the Lagrangian magnetic particle velocity are stretched exponentials (see Supp. Mat.~\cite{SuppMat}). However, we show that the large-scale modes are normally distributed by applying a spatial low-pass filter to the velocity field, confirming that the large-scale modes have reached a statistical equilibrium. The Kurtosis of the low-pass filtered velocity distribution is equal to 3 (inset Fig. \ref{fig03}). The shape of the PDF of the low-pass filtered velocities is also independent of the energy injection rate $\epsilon$, which is illustrated by the constant value of the kurtosis (inset Fig. \ref{fig03}). High-pass filtering of the velocity field also shows that the non-Gaussianity of the PDFs is reminiscent of the magnetic particle velocity one.

\begin{figure}[!t]
\begin{center}
\includegraphics[width=8.6cm]{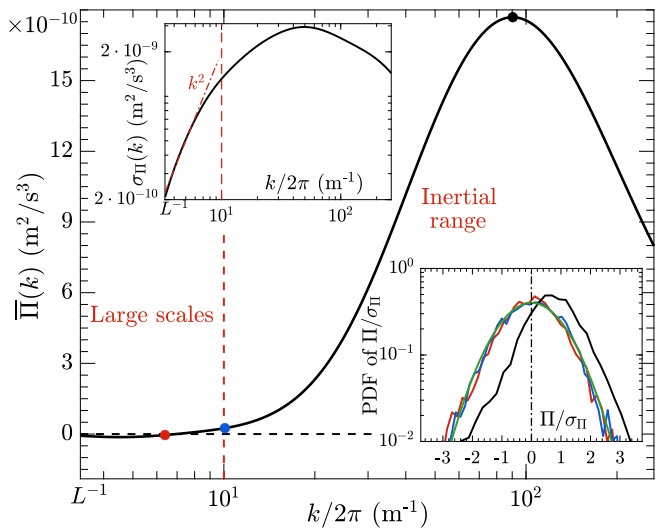} %fig_flux.eps
\caption{Time-averaged energy flux spectrum $\overline{\Pi}(k)$. At large scales ($k<k_i$), zero-mean energy flux is measured (equipartition). In the inertial range ($k>k_i$), the energy flux is positive and implies a direct cascade of energy. Insets: (top) Standard deviation of the energy flux spectrum $\sigma_{\Pi}(k)$. (bottom): PDFs of the fluctuations of the energy flux $\Pi/\sigma_{\Pi}$ for three values of the wavenumber $k$ (colored bullets). The green line represents a Gaussian distribution.} %$k=3.35$, 6.3, 10 and 90 m$^{-1}$.} %Gray lines show instantaneous spectrum, $\Pi(k,t)$.
\label{fig05}
\end{center}
\end{figure}

\paragraph*{Mean energy flux.\textemdash}Measuring the energy flux is essential to understanding the dynamics of turbulent phenomena~\cite{Verma}. We compute the time-averaged energy flux spectrum $\overline{\Pi}(k)$ from the expression $\Pi(k,t) = \langle \mathbf{v}^<_k \cdot [ \mathbf{v}^<_k \cdot \nabla \mathbf{v}^>_k]\rangle_{r} + \langle \mathbf{v}^<_k \cdot[ \mathbf{v}^>_k \cdot \nabla \mathbf{v}^>_k]\rangle_{r}$~\cite{Frisch}, where $\mathbf{v}^<_k(\mathbf{r}) \equiv \int_0^k\mathbf{\widehat{v}}(k')e^{i\mathbf{k'\cdot r}}\mathrm d k'$ is the low-filtered velocity field at the wavenumber $k$, $\mathbf{v}^>_k(\mathbf{r}) \equiv \int_k^\infty \widehat{\mathbf{v}}(k')e^{i\mathbf{k' \cdot r}}\mathrm d k'$ the high-filtered one, and $\mathbf{\widehat{v}}$ is the Fourier transform of the velocity field $\mathbf{v}$. The zero-mean energy flux measured at low wavenumbers confirms the statistical equilibrium regime (Fig.~\ref{fig05}). The interval in which the mean energy flux is zero corresponds to the same interval in which the $k^{2}$ power law of the energy spectrum is observed. In the inertial range, the energy flux is positive and implies a direct energy cascade towards high wavenumbers, corresponding to the turbulent cascade shown in Fig.~\ref{fig01}. For wavenumbers $k/2\pi$ higher than 100 m\textsuperscript{-1}, the energy flux strongly decreases, which is consistent with Fig.~\ref{fig01}. 

\paragraph*{Energy flux fluctuations.\textemdash}Although no energy cascades within the large scales in equipartition, intense temporal fluctuations of the energy flux $\Pi(k,t)$ are observed (bottom inset of Fig.~\ref{fig05} - see also Supp. Mat.~\cite{SuppMat}). This highlights that the large-scale domain is not isolated from the inertial range. Within the large-scale interval, the energy flux follows a Gaussian distribution (bottom inset of Fig.~\ref{fig05}), whose standard deviation $\sigma_{\Pi}(k) = \left[\overline{\Pi(k,t)^2}\right]^{1/2}$ is proportional to $ k^{2}$ (top inset), similarly to the energy spectrum $E(k)$. We can thus infer that $\sigma_{\Pi}(k)/E(k)\sim k^{0}$ for $k< k_i$. The damped fluctuations of zero-mean energy flux observed at low wavenumbers (top inset Fig.~\ref{fig05}) have also been reported in DNS~\cite{Dallas2015,AlexakisJFM2020}. The top inset of Fig.~\ref{fig05} also shows that these fluctuations are intense within the direct cascade but are strongly damped by viscous dissipation at high wavenumbers.

\begin{figure}[!t]
\begin{center}
\includegraphics[width=8.6cm]{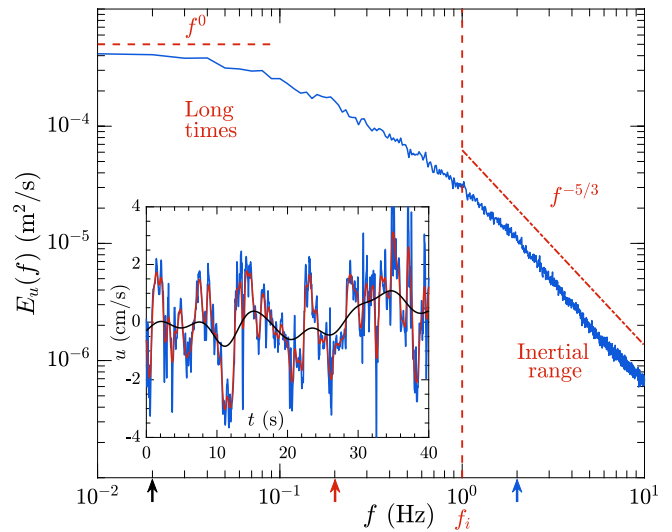} 
\caption{Temporal power spectrum density of the horizontal velocity $E_u(f)$. The dashed line represents a $f^{0}$ power law and the dot-dashed line represents a $f^{-5/3}$ power law. The forcing parameters are identical to Fig.~\ref{fig01}. $f_i$ indicates the beginning of the inertial range, i.e., the typical correlation frequency of the flow. Inset: Horizontal velocity $u(t)$ low-pass filtered at 2 Hz (blue), 0.2 Hz (red), and 0.02 Hz (black), as illustrated by the colored arrows in the main figure.} 
\label{fig04}
\end{center}
\end{figure}

\paragraph*{Temporal power spectrum.\textemdash}We now measure the temporal spectrum $E_u(f)$ of the horizontal velocity $u(t)$ (Fig.~\ref{fig04}). The signal is recorded for $\mathcal{T}=5$\, hours to converge the statistics at low frequencies ($f<f_i$), which represent the large-scale modes. To avoid a significant increase in the fluid temperature, we repeatedly performed LDV measurements for 100\,s and then we let the fluid cool down for 10\,min. The signal $u(t)$ is shown in the inset of Fig.~\ref{fig04}, with a low-pass filtered signal (black) to emphasize the slow modes of the temporal signal. The frequency spectrum shown in Fig.~\ref{fig04} is proportional to $f^{-5/3}$ at high frequencies ($f>f_i$). This is consistent with the $k^{-5/3}$ power law observed in the unidimensional energy spectrum, which implies a direct energy cascade in the inertial range (Fig.~\ref{fig01}). This power law was predicted by the Tennekes' model (large-scale advection of turbulent eddies) in isotropic turbulence without mean flow~\cite{Tennekes,CazaubielPRF2021}. At low frequencies ($f<f_i$), the frequency spectrum is found to be almost flat $f^{0}$, implying that large scales are uncorrelated. This is similar to the unidimensional spatial spectrum $E_{uu}(k)\sim k^0$ at large scales (Fig.~\ref{fig01}), suggesting that we observe the statistical equilibrium regime at low frequencies.

\paragraph*{Conclusion.\textemdash}We have experimentally shown that the large-scale dynamics in forced dissipative 3D hydrodynamic turbulence are in agreement with the statistical equilibrium prediction. This system is a remarkable example in which the large scales are in statistical equilibrium, while smaller scales are in an out-of-equilibrium stationary regime. 
A direct consequence of this experimental validation is that simulations leading to a statistical equilibrium regime, such as those of the truncated Euler equation~\cite{Dallas2015}, could provide a new tool to efficiently simulate the large-scale dynamics of 3D turbulent flows in various fields. Our findings also pave the way to possibly use concepts of equilibrium statistical mechanics (such as fluctuation-dissipation and fluctuation theorems) for large-scale turbulent flows. It can help better understand the interactions between the degrees of freedom at equilibrium (large scales) with out-of-equilibrium structures (small scales), which are essential when studying turbulent phenomena. In the future, we will explore the transient regimes of the equilibrium regime of large scales, called the thermalization processes, by measuring the growth and decay of turbulence~\cite{Saffman67,DavidsonJFM2012}. More generally, better identifying the mechanisms governing large-scale properties of turbulent flows such as statistical equilibrium, condensation, or inverse cascade, is of primary interest in 3D turbulence~\cite{Alexakis2018,Dallas2015,Xia2011,FalconPRF2017}, wave turbulence~\cite{Michel2017,FalconPRL2020,MiquelPRE2021}, and climate modeling~\cite{ShawARFM2013}.

\begin{acknowledgments}
We thank A. Cazaubiel, J.-C. Bacri, and S. Fauve for fruitful discussions. We thank P. Delor, A. Di Palma, Y. Le Goas, and V. Leroy for technical help. This work was supported by the French National Research Agency (ANR DYSTURB project No. ANR-17-CE30-0004) and by the Simons Foundation MPS N$^{\rm o}$651463-Wave Turbulence.
\end{acknowledgments}

%%%%%%%%%%%%%%%%%%%%%%%%%%%%%%%%%%%%%%
%%%%%%%%%%%% REFERENCES %%%%%%%%%%%%%%%%%%
%%%%%%%%%%%%%%%%%%%%%%%%%%%%%%%%%%%%%%


\begin{thebibliography}{99}
\bibitem{Note}In movies of Ref.~\cite{SuppMat}, note, in particular, the motions of the magnetic particles  are not synchronized with the vertical oscillating magnetic field.
\end{thebibliography}
\end{document}